\documentclass[aps,preprint]{revtex4}
\usepackage{epsfig}
\usepackage{amsmath}
\usepackage{epsfig}

\begin{document}
%\begin{twocolumn}
\title
{\bf Avoided crossings of energy levels in one-dimension}    
\author{Zafar Ahmed$^1$, Shashin Pavaskar$^2$, Dhruv Sharma$^3$, Lakshmi Prakash$^4$}
\affiliation{$^1$Nuclear Physics Division Bhabha Atomic Research Centre, Mumbai 400085, India\\ $^2$National Institute of Technology, Surathkal, Mangalore, 575025, India\\ $^3$National Institute of Technology, Rourkela, 769008, India \\ $^4$University of Texas, Austin, TX, 78705, USA}
\date{\today}
\begin{abstract}
\noindent
Due to the absence of degeneracy in one dimension, when a parameter, $\lambda$,  of a potential is varied 
slowly the discrete energy eigenvalue curves, $E_n(\lambda)$, cannot cross but they are allowed to come quite close
and diverge from each other. This phenomena is called avoided crossing of energy levels. The parametric evolution of eigenvalues of the generally known one dimensional potentials do not display avoided crossings and on the other hand some complicated and analytically unsolvable models do exhibit this.  Here, we show that this interesting spectral property can be found in simple one-dimensional double-wells when width of one of them is varied slowly.
\end{abstract}
\maketitle
In two or more dimensions, evolution of energy eigenvalues as one parameter ($\lambda$) of the system is varied slowly (adiabatically) may show two levels crossing (Fig. 1(a)) or coming quite close to each other and then diverging on either side of some special values of $ \lambda$ (see Fig. 1(b)). The former indicates degeneracy (equality of eigenvalues of two linearly independent  states of one Hamiltonian), the latter is called avoided crossing (AC). In a system, more frequent occurrence of AC  is debated to be the  signature of quantum Chaos [1] and non-integrability of the Hamiltonian (non-separability of a Hamiltonian, $H(x,y,z..)$, in various dimensions). For instance, the eigenvalues
of a particle in a two dimensional rectangular box as a function of the ratio of length to breath (L/B) show crossings  as its Hamiltonian is separable in $x$ and $y$. But if one small corner of this rectangle is snipped off, the eigenvalues show avoided crossings [2] because Hamiltonian is no more separable in $x$ $\&$ $y$ or in some other special co-ordinates: $\xi(x,y)$  $\&$ $\eta(x,y)$. Avoided crossings are known to be responsible for surprisingly low energy  Landau-Zener [3] transitions.

Crossings and avoided crossings of eigenvalues can  also be understood in terms of time independent perturbation theory at the textbook level. When a perturbation is switched on, the degeneracy of levels (when $\lambda=0$)  is lifted and they split into two or more levels (see Fig. 1(a)). Similarly, two un-perturbed levels $E_1$ and $E_2$ may split into sub-multiplets. Sub-multiplets of different levels may cross (see Fig 1(a)) each other. However, the sub-multiplets of one level have the tendency to avoid and diverge from each other as the parameter $\lambda$ is varied (see Fig. 1(b)). The Zeeman splitting [3] of different total angular momentum($J$) states under the magnetic field is one such example. The levels showing avoided crossing (Fig. 1(b)) can also called nearly-degenerate(nearly equal) levels and a perturbation widens the gap between them. 

On the other hand, the evolution of energy levels of one-dimensional potentials  under a slow change of the parameter is  generally  found to be uninteresting wherein all the levels increase/decrease monotonically without displaying avoided crossings  or crossings (due to absence of degeneracy). For instance see Fig. 1(c) for the evolution of first four energy eigenvalues of infinitely deep potential of width $\pi \lambda$.
In this regard, the sub-barrier( below the barrier) energy levels of a symmetric double-well [3] potential are  interesting which display the well known merging of two levels in to one as the height ($V_0$) or the width ($d$) of the in-barrier is increased. For moderate values of $V_0$ and $d$ the sub-barrier levels are closely lying doublets as in the classical case of Ammonia Molecule [3]. 

Presence of avoided crossings of energy levels in one-dimensional potentials has been addressed once [5] in extremely complex situations wherein the Schr{\"o}dinger equation becomes Heun equation which can not be solved exactly and analytically. However elegant approximate treatments have been employed to display avoided crossings of eigenvalues even in one dimension. A recent book [6] is a rare avenue where the avoided crossings are well discussed and exhibited  in  piecewise constant one dimensional asymmetric  double-well potentials of finite support in terms of inherent transmission resonances. These potentials of finite support represent open systems that can have all: scattering, bound and resonant states.

Here, we present various analytically solvable asymmetric double well potentials  with a fixed in-barrier. We show that these closed systems which support only bound states exhibit avoided crossing in one dimension when the width of one of the wells is varied slowly. 
\begin{figure}
\centering
\includegraphics[width=5 cm,height=3.5 cm]{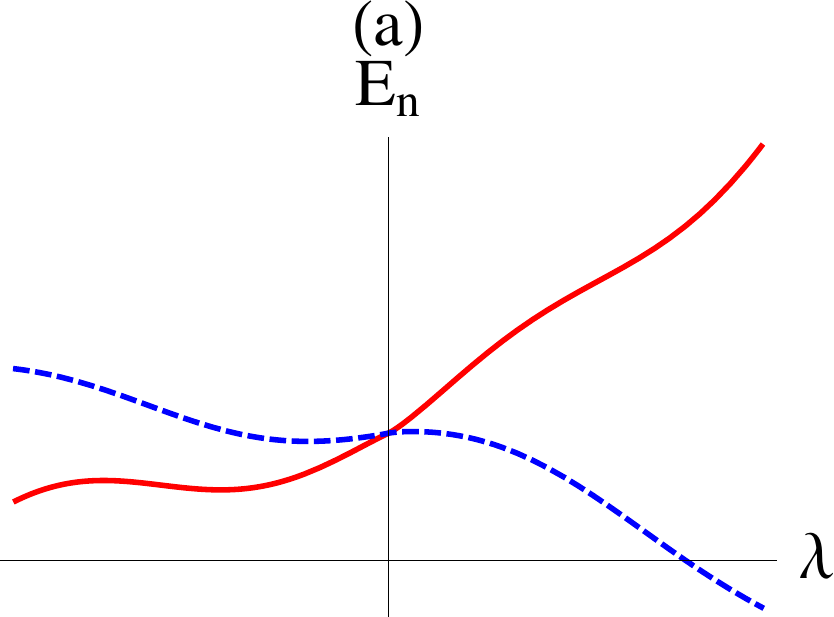}
\hskip .5 cm
\includegraphics[width=5 cm,height=3.5 cm]{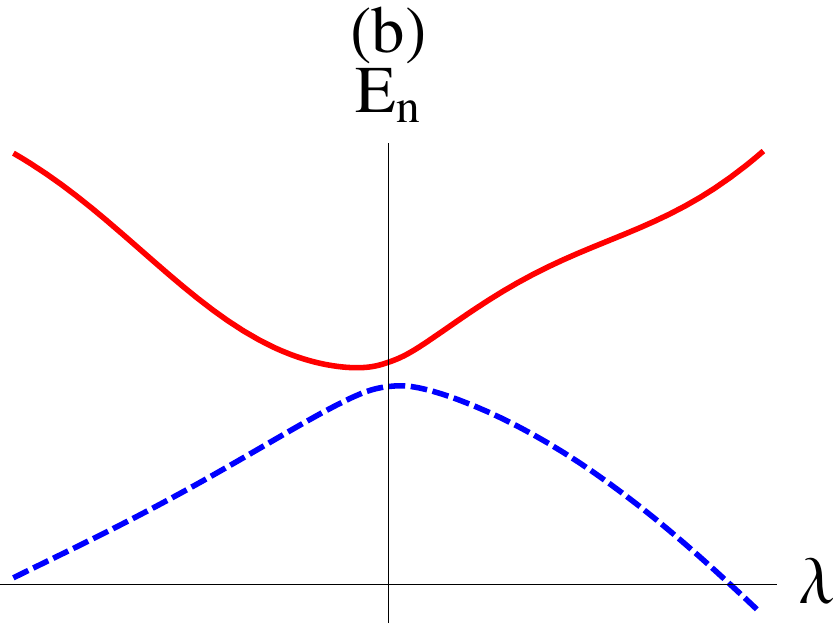}
\hskip .5 cm
\includegraphics[width=5 cm,height=3.5 cm]{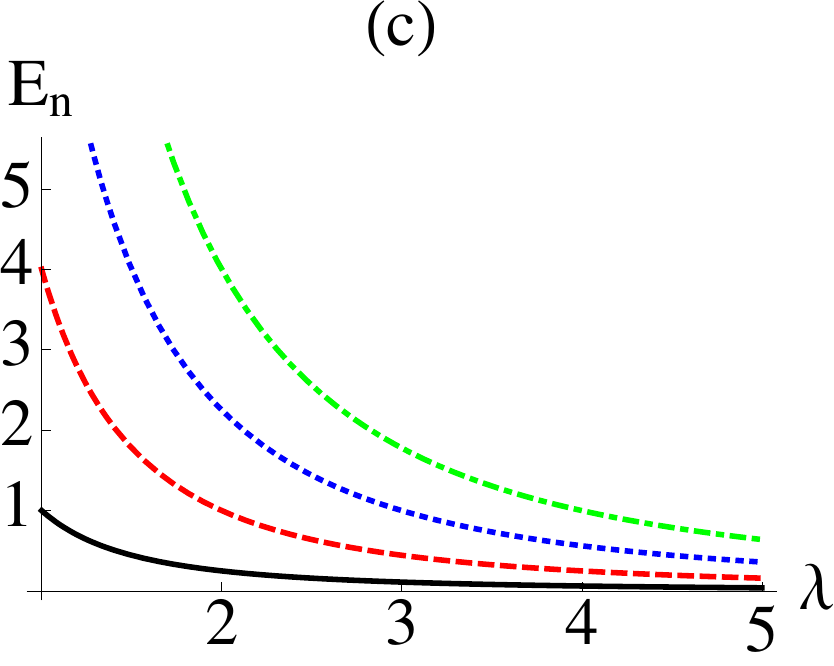}
\caption{Schematic evolution of two eigenvalues depicting (a): crossing of two levels (degeneracy), (b): avoided crossing (level repulsion) , 
(c): evolution of energy levels of  one dimensional deep infinite well with parameter $\lambda$: $E_n=(n/\lambda)^2$. Notice that there is no crossing or avoided crossing of two levels.}
\end{figure}

In quantum mechanics, we solve the Schr{\"o}dinger equation: $H(x) \psi_\alpha(x)=E_\alpha \psi_\alpha(x)$, there may or may not be a one to one correspondence between energy eigenvalues $E_\alpha$ and $\psi_\alpha(x)$. For example for free particle case when $H=-\frac{\hbar^2}{2m}\frac{d^2}{dx^2}$, for one real positive energy $E$, there are two eigenstates $e^{ikx}$ and $e^{-ikx}$, where $k =\frac{\sqrt{2mE}}{\hbar}$.
Similarly in the presence of a  potential, $V(x)$ for real positive energy continuum there are two states $\psi_{k}(x)$ and $\psi_{-k}(x)$.

However in one dimension for bound states of a potential this correspondence is strictly one to one as an eigenstate has to  satisfy the Dirichlet boundary condition: $\psi(\pm \infty)=0$. Interestingly, the Schr{\"o}dinger equation being second order differential equation 
\begin{equation}
\frac{d^2\psi(x)}{dx^2}+\frac{2\mu}{\hbar^2}[E-V(x)]\psi(x)=0
\end{equation}
there exist two linearly independent solutions. However, usually only one of them satisfies the said boundary condition. Further, it can be proved that even if there are two solutions: $\psi_m(x)$ and $\psi_n(x)$ for one fixed discrete energy then they are only trivially different as they would be linearly dependent (see Appendix 1).

\begin{figure}[ht]
\centering
\includegraphics[width=3.5 cm,height=5 cm]{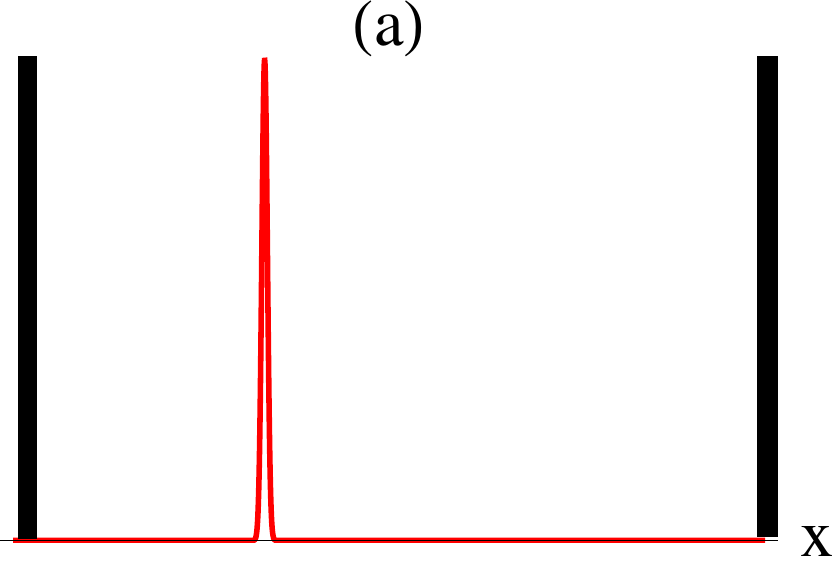}
\hskip .5 cm
\includegraphics[width=3.5 cm,height=5 cm]{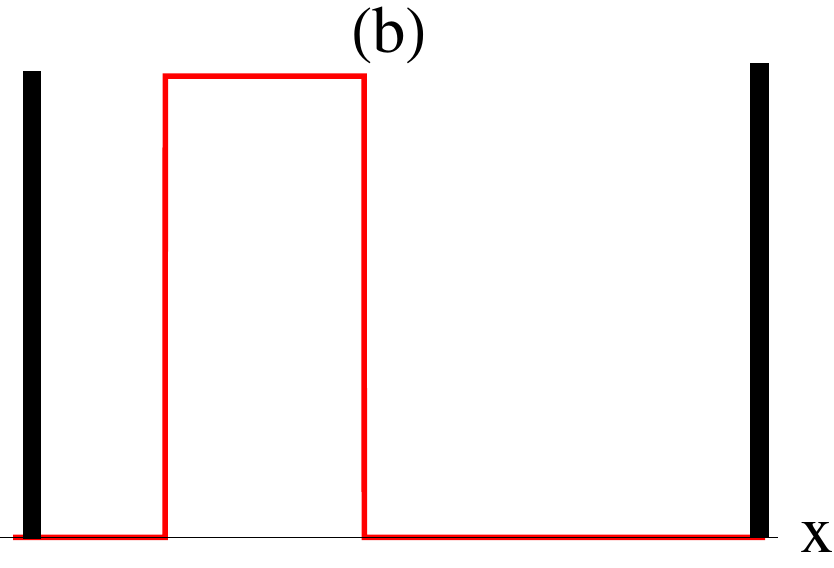}
\hskip .5 cm
\includegraphics[width=3.5 cm,height=5 cm]{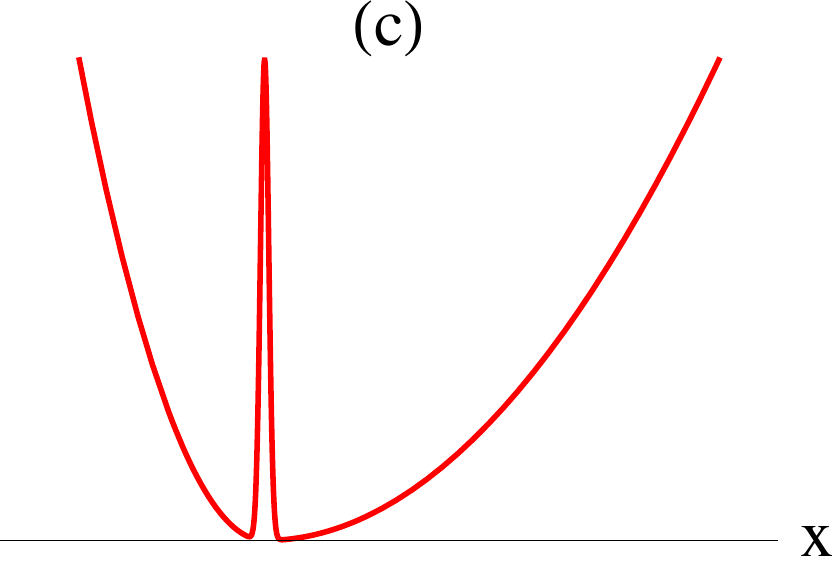}
\hskip .5 cm
\includegraphics[width=3.5 cm,height=5 cm]{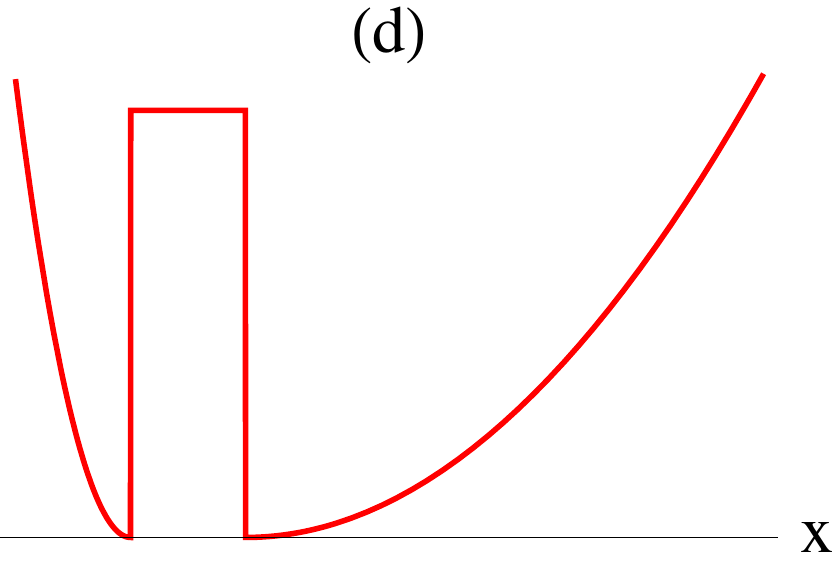}
\caption{Depiction of various double well potentials. (a) M1 Eq. (2), (b) M2 Eq. (7), (c) M3 Eq. (14) and (d) M4 Eq. (19). In the asymmetric harmonic wells the value of the parameter $\hbar\omega$ on the left is more than that of right}
\end{figure}

Interestingly, it requires two or more dimensions for degeneracy to occur where energy eigenvalues for separable potentials is a function of two positive integers $n_1$ and $n_2$. For instance in the case of particle in a two dimensional square box of length $\pi$ ($2\mu=1=\hbar^2$), we have  i.e., $E_{n_1,n_2}=n_1^2+n_2^2$, where $E=260$ is four-fold degenerate as $2^2+16^2=260=8^2+14^2$ the corresponding eigenstates are $\psi_1(x,y)=A \sin{2x}~\sin{16y}$ and $\psi_2(x,y)=A \sin 8x ~ \sin 14 y$. This is called accidental degeneracy. Additional degeneracies arise from the symmetries of Hamiltonian. For example, we can have two more eigenstates due to the symmetry of Hamiltonian under $x \leftrightarrow y$  as 
$\psi_3(x,y)=A \sin 2y~ \sin 16x$ and $\psi_4(x,y)= A \sin 8y ~ \sin 14x$. A two dimensional Hamiltonian
may not be (apparently) separable in $x$ and $y$ then quantum numbers $n_1$ and $n_2$ loose meaning and eigenvalues are found  by semi-classical methods like periodic orbit theory [1]. 

In more than one dimensions when eigenvalues are studied by varying a parameter smoothly and slowly,
$E_{n_1,n_2}(\lambda)$ evolve and may become wavy curves rising up and lowering down having crossings and avoided crossings. The eigenvalues of anisotropic  harmonic oscillator as a function of anisotropy parameter is a simple example where eigenvalue curves are slant straight lines [4]. Nilsen's [4] modified anisotropic Harmonic oscillator model of deformed nuclei displays all raising and lowering, crossings and avoided crossings of energy levels as a function of the  deformation parameter. Eventually, this explains the observed nuclear magic numbers.  The similar phenomena are also found in quantum dots.

It is known that the first few energy levels of a double well potential are closely lying doublets below or around the height of the in-barrier.  We bring out a  scenario wherein the wells of the same depth but different width display avoided crossings  when the width of one of the wells is varied slowly (adiabatically). Suppose in a symmetric double well potential, we have two very close energy eigenvalues, when the width parameter is increased or reduced slightly these two levels would diverge from each other to display avoided crossing, what is counter intuitive is the occurrence of AC  when the asymmetry is large. This is another interesting  manifestation of Sch{\"o}dinger equation  and of quantum mechanics.

In the following, we find the eigenvalue formulae for four double well potentials depicted in Fig. 2. We define $u=2\mu/\hbar^2$ and take $u=1~ (eV {A^0}^2)^{-1}$ (in Figs. 3,5,6) which corresponds to $\mu \approx 4 m_e$, where $m_e$ is the mass of electron in $eV$, the energy is measured in $eV$ (electron volts)and the lengths $a,b$ and $c$ are measured in Angstrom $(A^0)$. In the case of Fig. 4, we have taken take $\mu= m_e$ so $u=0.2625~ (eV {A^0}^2)^{-1}$.

\noindent
{\bf M1: Dirac Delta between two rigid walls}

This potential is given as
\begin{equation}
V(x)=V_0 \delta(x)+\left \{\begin{array}{lcr}
0, & & -a<x<b,\\
\infty & & x<-a,x>b
\end{array}
\right.
\end{equation}
This potential is depicted schematically in Fig. 2(a).
When Schr{\"o}dinger equation has Dirac Delta potential at $x=0$,  the first derivative of the wave function becomes discontinuous at $x=0$ as [7]
\begin{equation}
\psi'(x>0)-\psi'(x<0)=(2mV_0/\hbar^2) \psi(0)
\end{equation}
The appropriate solution of the Schr{\"o}dinger equation for this double well
potential can be written as
\begin{equation}
\psi(x)= \left\{\begin{array}{lcr}
A \sin k(x+a), & & x\ge -a,\\
B \sin k(x-b) & & x\le b,
\end{array}
\right.
\end{equation}
satisfying
the Dirichlet boundary condition that $\psi(-a)=0=\psi(b)$. The continuity
and the mismatch condition (3) gives
\begin{eqnarray}
A \sin ka= -B \sin kb,\\ \nonumber
Bk \cos kb - Ak \cos ka =(2\mu /\hbar^2) V_0 A \sin ka.
\end{eqnarray}
The elimination of $A,B$ gives us the eigenvalue formula for  bound states of the potential (2) as
\begin{equation}
\sin k(a+b)=- (2mV_0/\hbar^2) \sin ka \sin kb
\end{equation}
When $V_0=0$, we get $k_n(a+b)=n\pi$ the energy eigenvalues of infinitely deep well of width $a+b$. When $a=b$ in (6), we get the usual eigenvalue formula for the symmetric case as $k \cot ka=-(mV_0/\hbar^2)$ [8]. We take $u=1 (eV {A^0}^2)^{-1}, V_0=10 eV, a=2 A^0$ and vary $b$ in $[0-5] A^0$ in Eq.(6) to calculate  and show the first four eigenvalues in Fig. 3(a). Notice ACs at $ b\sim 1,2, 4 (A^0)$. The first two ACs are very clear in the enlarged plots 3(b,c).

In Fig. 4(a), we present the first four eigenvalues $E_n$ for M1 for $u=0.2625 (eV {A^0}^2)^{-1}, V_0= 20 eV, a=5 A^0$.  We vary $b$ in $[0,25] A^0$. Avoided crossings can be seen for $E \sim 5,10,15 (eV)$. In part (b) we plot $E'_n=u(a+b)^2 E_n$ to notice the interesting wavy evolution of eigenvalues  exhibiting ACs clearly. This evolution is similar to the case of Fig. 28.2 and 28.3 of Ref. [6] which have been obtained for double wells of finite support and discussed at par with the real part of the resonant energies. 

\noindent
{\bf M2: Asymmetric rectangular double well potential}

The simple (symmetric) double well potential is an often-discussed problem in textbooks [9,10]. However here we need to discuss the asymmetric double well
\begin{equation} 
V(x)=\left\{\begin{array}{lcr}
\infty, & & x\le -a, x\ge c\\
0, & & -a <x<-b, b<x<c,\\
V_0, & & -b\le x\le b,\\
\end{array} 
\right.
\end{equation}
and obtain the formula for discrete energy eigenvalues. We write the solution of (1) for this potential as
\begin{equation}
\begin{array}{lcr}
\psi(x)=A \sin k(x+a),& & x \in [-a,-b]\\
\psi(x)=D \sin k(x-c),& & x \in [b,c].
\end{array}
\end{equation}
For the region $-b<x<b$ we have
\begin{equation}
\psi(x)=B \sinh p x + C \cosh p x,\quad \mbox{where} \quad p=\frac{\sqrt{2m(V_0-E)}}{\hbar^2}. 
\end{equation}
We match the solutions and their derivative at $x= -b$, we get
\begin{eqnarray}
A \sin k(a-b) = -B \sinh p b + C \cosh p b\\ \nonumber
k A \cos k(a-b)= p B \cosh p b-p C \sinh p b
\end{eqnarray}
Similarly, the matching conditions at $x=b$ give
\begin{eqnarray}
 B \sinh p b + C \cosh p b= D \sin k(b-c)\\ \nonumber
p B \cosh p b+p C \sinh p b =D k\cos k(b-c)
\end{eqnarray} 
We now demand the consistency of the above four Eqs. (10,11) and for their non-trivial (non-zero) solutions for $A,B,C,D$, we get 
\begin{equation}
\left |\begin{array} {cccc} -\sinh pb & 
\cosh pb & 0 & \sin kd_1  \\ p \cosh pb & -p \sinh p b & 0 & k \cos kd_1 \\ \sinh pb & \cosh p b & \sin kd_2 & 0 \\  p \cosh pb & p \sinh pb & -k \cos k d_2 & 0
\end{array} \right |=0.
\end{equation} 
Opening the determinant (12) and introducing   $a-b=d_1, c-b=d_2, d=d_1+d_2$, we get
\begin{equation}
\begin{array}{c}
\sqrt{E(V_0-E)}~ \cosh 2pb ~\sin k d + [E \cos kd + V_0 \sin kd_1~ \sin kd_2]~ \sinh 2bp=0, \\
\end{array}
\end{equation}
when $E \ne 0, V_0$ as these cases require separate(special) linear solution $\psi(x)= Ax+b$ [11] in the region where $E-V(x)$ is zero.

In Fig. 5(a), we calculate and show the first five eigenvalues of this model (7) for $u=1~(eV {A^0}^2)^{-1}, V_0= 10 eV, a= 2 A^0, b=1 A^0$  from  Eq. (13). The parameter $c$ is varied smoothly in $[1,6] A^0$. Notice very interesting ACs about the horizontal line $E \sim 5.37 eV$ at $E \sim 2.1, 3.4, 5, 6 (eV)$ which show clearly in the enlarged plot 5(b).
\begin{figure}
\centering
\includegraphics[width=5 cm,height=5 cm]{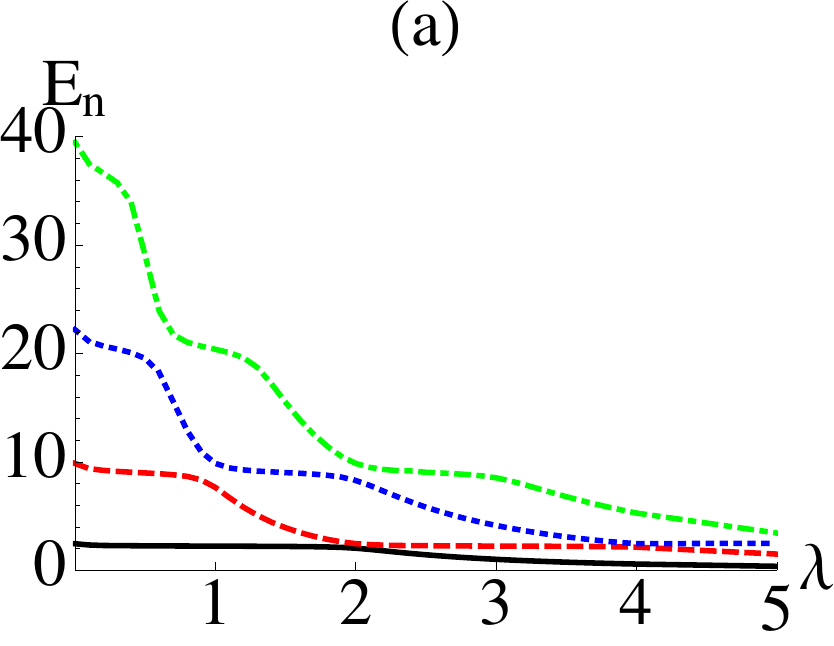}
\hskip .5 cm
\includegraphics[width=5 cm,height=5 cm]{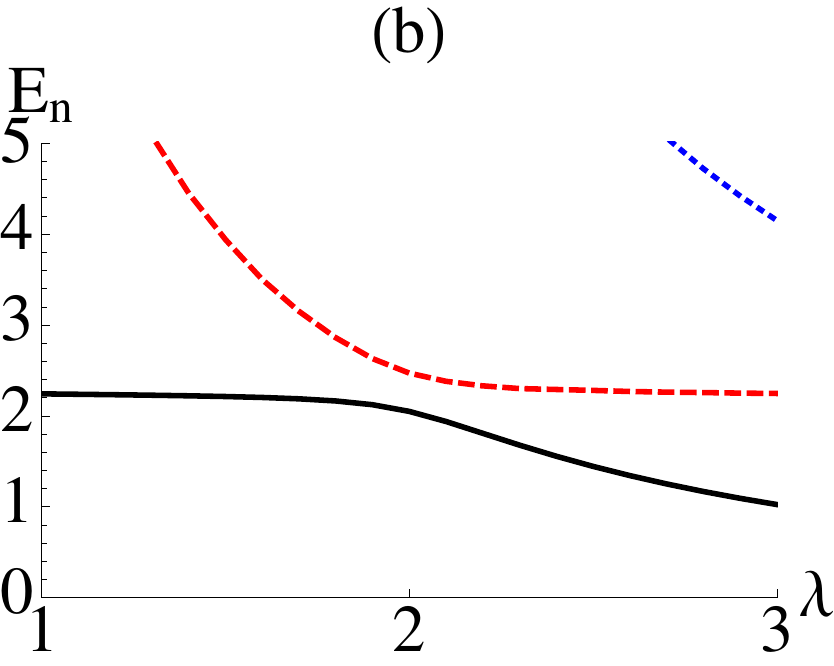}
\hskip .5 cm
\includegraphics[width=5 cm,height=5 cm]{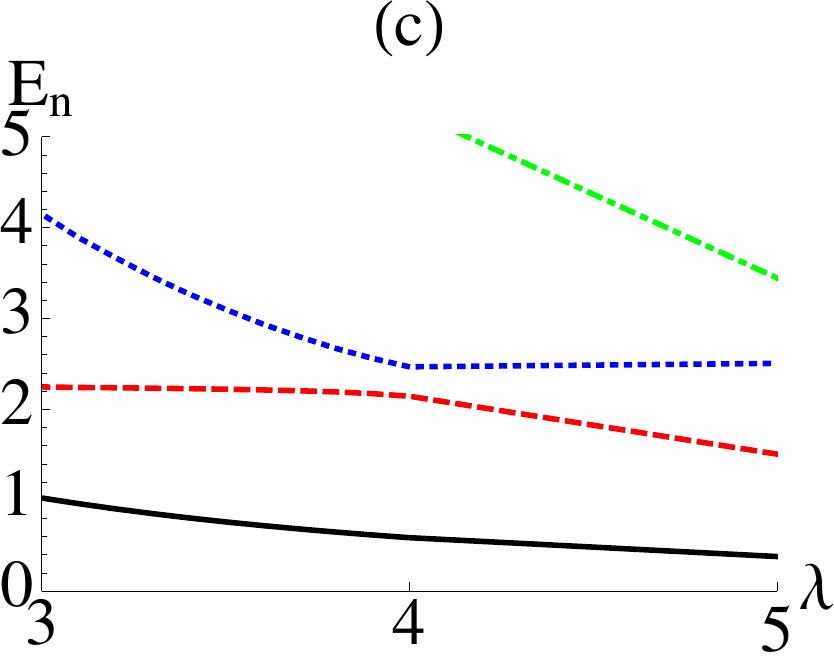}
\caption{(a) Evolution of first four discrete eigenvalues of the model M1.  $V_0=10 eV, a=2 A^0$ and $b$ is varied in $[0,5]$. In (a) for $ a \le 1 A^0$ two avoided crossing can be seen clearly. Around  the symmetric case $(a=2 A^0=b)$ there are two avoided crossings.  In the enlarged parts  (b) and (c) the first two ACs appear clearly. Here $\lambda= b, \lambda_s=2 A^0$}
\end{figure}

\noindent
{\bf M3: Dirac Delta in asymmetric Harmonic well}

This potential can be expressed as 
\begin{equation}
V(x)=V_0 \delta(x)+\left\{ \begin{array}{lcr}
\frac{1}{2}\mu \omega_1^2 x^2, & & x< 0,\\
\frac{1}{2}\mu \omega_2^2 x^2,, & & x>0
\end{array}
\right.
\end{equation}
For the harmonic oscillator potential $V(x)=\frac{1}{2}\mu\omega^2 x^2$
we introduce $z=bx,\alpha=\sqrt{2\mu \omega/\hbar}$ and $\nu=E/(\hbar\omega)-1/2$ to write (1) as
\begin{equation}
\frac{d^2\psi(z)}{dz^2}+[\nu+1/2-z^2/4]\psi(z)=0, z=bx, \alpha=\sqrt{2\mu \omega/\hbar}, \nu=E/(\hbar \omega)-1/2.
\end{equation}
This equation admits two linearly independent solutions $D_{\nu}(z), D_{\nu}(-z)$ called parabolic cylindrical functions. As $z=\rightarrow \infty$, $D_{\nu}(z)$ tends to zero. So in order to satisfy Dirichlet boundary condition: $\psi(\pm \infty)=0$, we choose 
\begin{equation}
\psi(x)=\left\{ \begin{array}{lcr}
A D_{\nu_1}(-\alpha_1 x), & & x< 0,\\
B D_{\nu_2}(\alpha_2 x)& & x>0
\end{array} \quad  z_i=\alpha_i x, \alpha_i=\sqrt{2\mu \omega_i/\hbar}, i=1,2.
\right.
\end{equation}
By matching these solutions and mismatching (3) their derivative at $x=0$, we
get the energy eigenvalue equation as
\begin{equation}
\sqrt{\hbar \omega_2}~ \frac{D'_{\nu_1}(0)}{D_{\nu_2}(0)} +\sqrt{\hbar \omega_1}~ \frac{D'_{\nu_1}(0)}{D_{\nu_1}(0)}= V_0\sqrt{2\mu/\hbar^2}
\end{equation}
which simplifies to
\begin{equation}
\sqrt{2\hbar \omega_2}~ \frac{\Gamma[1/2-\nu_2/2]}{\Gamma[-\nu_2/2]}+\sqrt{2\hbar \omega_1}~ \frac{\Gamma[1/2-\nu_1/2]}{\Gamma[-\nu_1/2]}=-V_0\sqrt{2\mu/\hbar^2},
\end{equation}
by using the analytic expressions of $D_{\nu}(0)$ and $D'_{\nu}(0)$ [12].
In Fig. 6(a) we present the evolution of first five eigenvalues of the model M3 using (18) for $u=1~(eV {A^0}^2)^{-1}, V_0 = 10 eV, \hbar\omega_1=2 eV$ by varying $\hbar\omega_2$ in $(0,3] eV$. Notice two ACs below $\hbar \omega_2=1 eV$, one above $E=1 eV$ and two ACs around $\hbar\omega_2=2 eV$ (at this value the double well potential (14) is symmetric).

\noindent
{\bf M4: Rectangular barrier in asymmetric harmonic well}
\begin{equation} 
V(x)=\left\{\begin{array}{lcr}
\frac{1}{2}\mu \omega_1^2(x+a)^2, & & x\le -a,\\
V_0, & & -a <x<a,\\
\frac{1}{2}\mu \omega_2^2 (x-a)^2, & & \ge a,\\
\end{array}
\right.
\end{equation}
The solution of (1) for (19) can be written as
\begin{equation} 
\psi(x)=\left\{\begin{array}{lcr}
F D_{\nu_1}[\alpha_1(-x-a)], & & x\le -a, \\
A \sinh qx + B \cosh qx, & & -a <x<a,\\
G D_{\nu_2}[\alpha_2(x-a)], & & x\ge a,\\
\end{array}
\right.
\end{equation}
Matching the wave function and its derivative at $x=-a$, we get
\begin{eqnarray}
-A \sinh qa + B \cosh qa = F D_{\nu_1}(0),\\ \nonumber
q A \cosh qa - q B \sinh qa= -\alpha_1 F D_{\nu_1}'(0) 
\end{eqnarray}
and at $x=a$, we get
\begin{eqnarray}
A \sinh qa + B \cosh qa = G D_{\nu_2}(0), \\ \nonumber
qA \cosh qa + qB \sinh qa = \alpha_2 G D'_{\nu_2}(0).
\end{eqnarray}

Finding $A/B$ from Eqs.(21,22) equating them and then using the expressions for
$D_{\nu}(0)$ and $D'_{\nu}(0)$, we obtain
\begin{eqnarray}
\left(\sqrt{2\hbar \omega_1}~ \frac{\Gamma[1/2-\nu_1/2]}{\Gamma[-\nu_1/2]}+\sqrt{2\hbar \omega_2}~ \frac{\Gamma[1/2-\nu_2/2]}{\Gamma[-\nu_2/2]}\right) q\cosh 2 qa= \\ \nonumber \left (q^2 + 2\sqrt{\hbar\omega_1 \hbar \omega_2}
\frac{\Gamma[1/2-\nu_1/2]\Gamma[1/2-\nu_2/2]}{\Gamma[-\nu_1/2]\Gamma[-\nu_2/2]}\right) \sinh 2qa, \quad q=\frac{\sqrt{2\mu(V_0-E)}}{\hbar^2}.
\end{eqnarray}
Using this Eq. (23), we calculate the evolution of the first five eigenvalues $E_n(\lambda)$ of the model M4 for $u=1~(eV {A^0}^2)^{-1}, V_0=10 eV,\hbar\omega_1=2 eV$ and by varying $\lambda=\hbar\omega_2$ in $(0,3]  eV$. Fig. 6(b) shows straight lines having two ACs below $\hbar \omega_2=1 eV$, one above $E =1 (eV)$ and two around the symmetric case $\hbar \omega_2=2 (eV)$. The straight line behaviour of $E_n(\lambda)$ is similar to the one obtained in Figs. 1,2 of Ref.[5] where the Schr{\"o}dinger equation became Heun's second order differential equation.
\begin{figure}
\centering
\includegraphics[width=7.5 cm,height=5 cm]{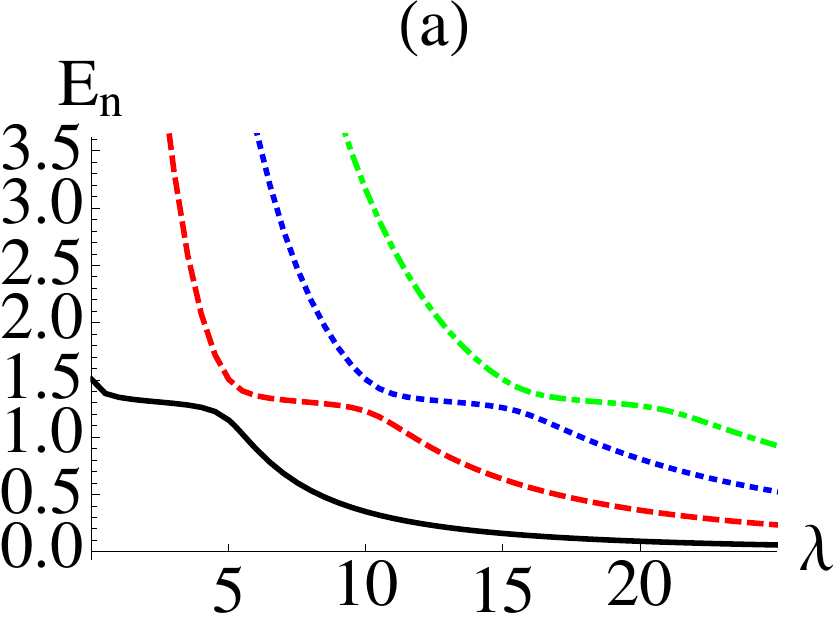}
\hskip .5 cm
\includegraphics[width=7.5 cm,height=5 cm]{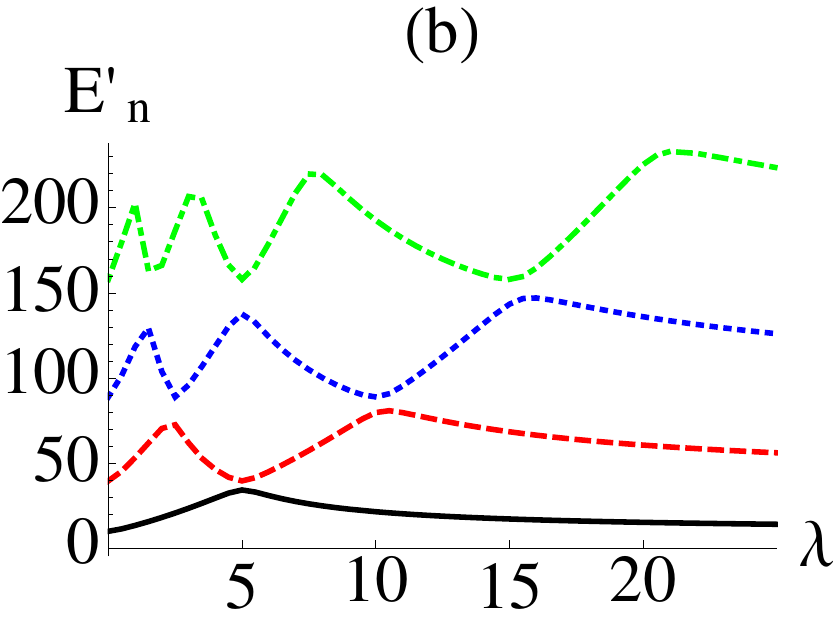}
\caption{First four eigenvalues ($E_n(\lambda)$) of the model M1 in (a) and effective eigenvalues  $E'_n=2\mu(a+b)^2E_n/\hbar^2$ in (b). We take $\mu=m_e, u=0.2625~ (eV {A^0}^2)^{-1}, V_0=20 eV, a=5 A^0$, $b$ is varied in $[0,25] A^0$. Here gain $\lambda =b, \lambda_s= 5$.}
\end{figure}

\begin{figure}
\centering
\includegraphics[width=7.5 cm,height=5 cm]{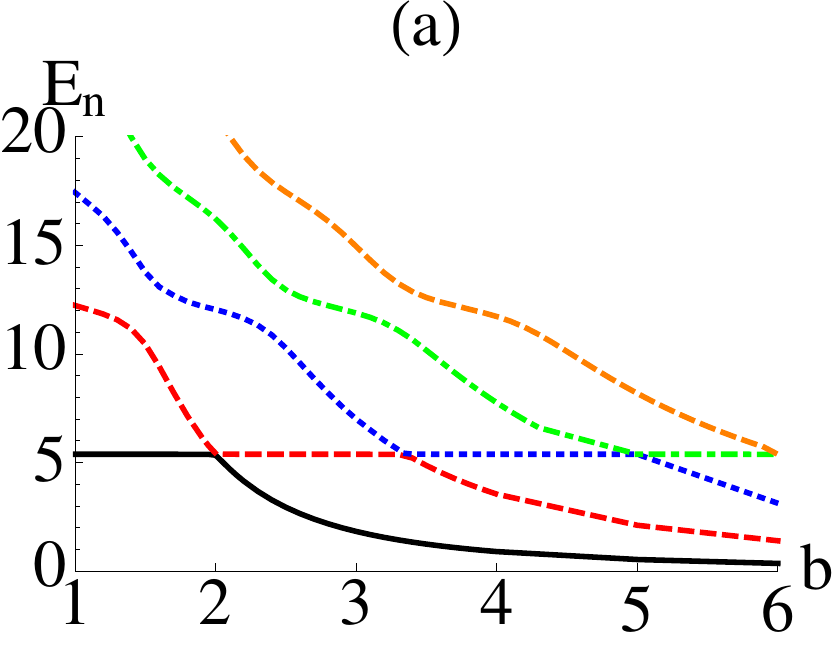}
\hskip .5 cm
\includegraphics[width=7.5 cm,height=5 cm]{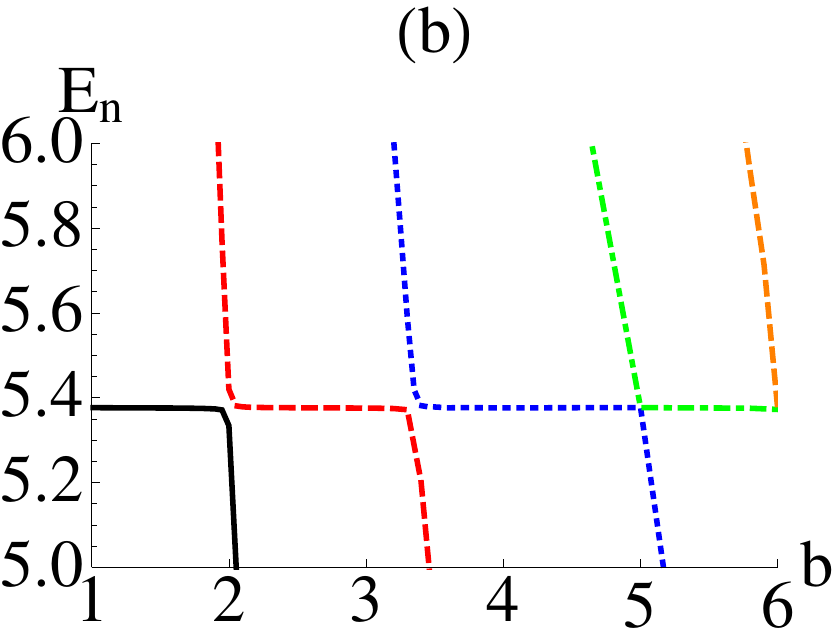}
\caption{(a) Evolution of first five discrete eigenvalues of rectangular double well potential (7). We take $V_0=10 eV$, $a=2 A^0, b=1 A^0$, and the width of the right well is being varied by changing the value of $c$ in [1, 6]. In (a) it appears that the straight line at $E \sim 5.37 eV$ is crossing other three curves around $c \sim 2.1, 3.4, 5, 6$. But the difference in the types  of curves above and below this line at  negate crossings. In the enlarged figure (b) see the three avoided crossings around  $c=2.1,3.4,5, 6$. Here $\lambda=c, \lambda_s=2 A^0$.}
\end{figure}

Based on our investigations of four double well potentials of various kinds, presented here, we conclude that when width ($\lambda$) of one of the wells is varied slowly the avoided crossings of eigenvalues are observed. With the so far understanding of double well potentials, the  ACs around the point of symmetry $(\lambda=\lambda_s)$ may be expected. Nevertheless, the ACs much below or much above this point are counter intuitive and they invite attention for further investigations.

\begin{figure}
\centering
\includegraphics[width=7.5 cm,height=5 cm]{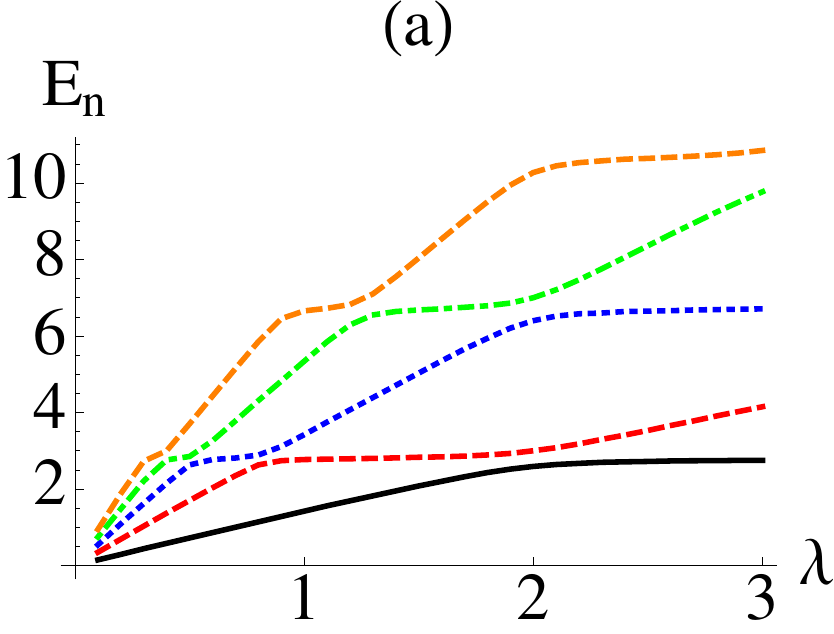}
\hskip .5 cm
\includegraphics[width=7.5 cm,height=5 cm]{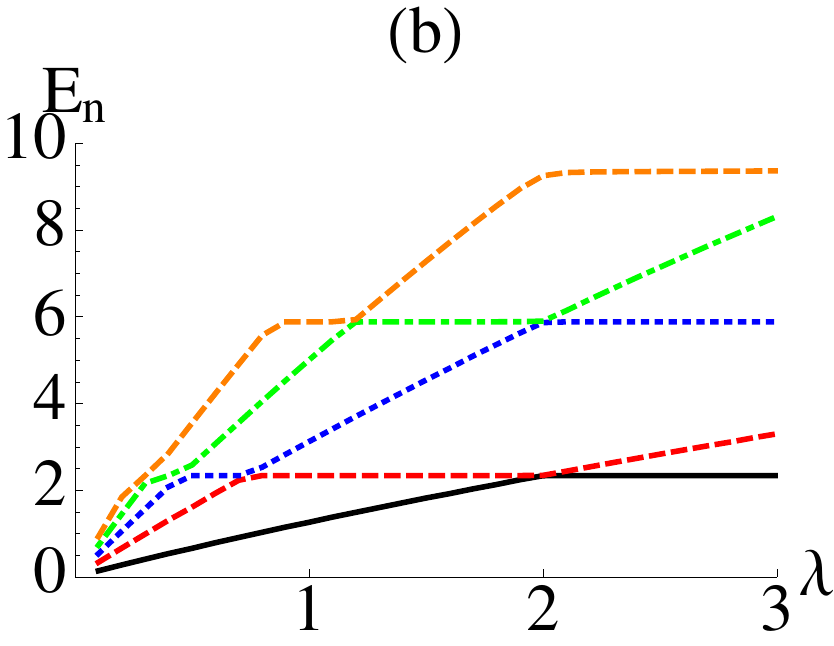}
\caption{First five eigenvalues of the models M3 and M4 when $V_0=10 eV$ in (a) and (b), respectively.  The  parameter $\hbar \omega_1= 2 eV$ for the left well and vary $\hbar \omega_2$ in $(0,3]$ for the right well. Notice two ACs below  $\hbar \omega_2 = 1 eV$, one above this and two around  $\hbar\omega_2=1 eV$. Here $\lambda=\hbar\omega_2$ and $\lambda_s= 2 eV$}
\end{figure}
\section*{\Large{Appendix 1}}
\renewcommand{\theequation}{A-\arabic{equation}}
\setcounter{equation}{0}
\noindent
{\bf Proposition 1:}
Let $\psi_m(x), \psi_n(x)$ be two $L^2-$integrable solutions of one dimensional time-independent Schr{\"o}dinger equation satisfying Dirichlet boundary
condition: $\psi(\pm \infty)$ with an equal eigenvalue $E$, then $\psi_m(x)$ and $\psi_n(x)$ are linearly dependent. 

{\bf Proof:}
Let the potential $V(x)$ (real or complex) in Schr{\"o}dinger $(2\mu =1=\hbar^2)$ equation gives rise to two solutions
$\psi_m(x)$ and $\psi_n(x)$ with the same energy eigenvalue $E$, then we write
\label{allequations}
\begin{eqnarray}
\frac{d^2\psi_m(x)}{dx^2}+[E-V(x)]\psi_m(x)=0, \\ 
\frac{d^2\psi_n(x)}{dx^2}+[E-V(x)]\psi_n(x)=0.
\end{eqnarray}
Multiply the first by $\psi_n(x)$ and the second by $\psi_m(x)$  and by subtracting them we get
\begin{equation}
\psi_m(x) \frac{d^2\psi_n(x)}{dx^2}-\psi_n(x) \frac{d^2\psi_m(x)}{dx^2}= 0 \Rightarrow \frac{d}{dx}\left (\psi_m \frac{d\psi_n}{dx}-
\psi_n \frac{d\psi_m}{dx}\right)=0,
\end{equation}
leading to
\begin{equation}
\left ( \psi_m(x) \frac{d \psi_n(x)}{dx}-
\psi_n(x) \frac{d \psi_m(x)}{dx} \right) = C,
\end{equation}
where $C$ is constant independent of $x$ which can as well be determined at $x=\pm \infty$. As the eigenstates satisfy $\psi_j(\pm \infty)=0$, we get $C=0$. Further we get, $\frac{\psi_m'(x)}{\psi_m} = \frac{\psi_n'(x)}{\psi_n}$ implying linear independence: $\psi_m(x)= C' \psi_n(x)$.

\end{document}